\documentclass[twocolumn]{aastex62}

\usepackage{lineno}
\usepackage{apjfonts}
\usepackage{subfigure}
\usepackage{amsmath}
\usepackage{xspace}
\usepackage{hyperref}
\usepackage[flushleft]{threeparttable}
\usepackage{multirow}

\newcommand{\htwoo}{H$_2$O\xspace}
\newcommand{\cotwo}{CO$_2$\xspace}

\shorttitle{Gas giant formation histories from refractory content}
\shortauthors{Chachan et al.}

\begin{document}

\title{\textbf{\large{Breaking Degeneracies in Formation Histories by Measuring Refractory Content in Gas Giants}}}
\correspondingauthor{Yayaati Chachan}
\email{yayaati.chachan@mcgill.ca}

\author[0000-0003-1728-8269]{Yayaati Chachan}
\altaffiliation{CITA National Fellow}
\affil{Department of Physics, McGill University, 3600 Rue University, Montreal, QC H3A 2T8, Canada}
\affil{Division of Geological and Planetary Sciences, California Institute of Technology, 1200 E California Blvd, Pasadena, CA 91125, USA}

\author[0000-0002-5375-4725]{Heather A. Knutson}
\affil{Division of Geological and Planetary Sciences, California Institute of Technology, 1200 E California Blvd, Pasadena, CA 91125, USA}

\author[0000-0002-5375-4725]{Joshua Lothringer}
\affil{Department of Physics, Utah Valley University, 800 W. University Pkwy., Orem, UT 84058, USA}

\author[0000-0002-5375-4725]{Geoffrey A. Blake}
\affil{Division of Geological and Planetary Sciences, California Institute of Technology, 1200 E California Blvd, Pasadena, CA 91125, USA}

\begin{abstract}

Relating planet formation to atmospheric composition has been a long-standing goal of the planetary science community. So far, most modeling studies have focused on predicting the enrichment of heavy elements and the C/O ratio in giant planet atmospheres. Although this framework provides useful constraints on the potential formation locations of gas giant exoplanets, carbon and oxygen measurements alone are not enough to determine where a given gas giant planet originated. Here, we show that characterizing the abundances of refractory elements (e.g., silicon, iron) can break these degeneracies. Refractory elements are present in the solid phase throughout most of the disk and their atmospheric abundances therefore reflect the solid-to-gas accretion ratio during formation. We introduce a new framework that parameterizes the atmospheric abundances of gas giant exoplanets in the form of three ratios: Si/H, O/Si, and C/Si. Si/H traces the solid-to-gas accretion ratio of a planet and is loosely equivalent to earlier notions of ‘metallicity’. For O/Si and C/Si, we present a global picture of their variation with distance and time based on what we know from the solar system meteorites and an updated understanding of the variations of thermal processing within protoplanetary disks. We show that ultra-hot Jupiters are ideal targets for atmospheric characterization studies using this framework, as we can measure the abundances of refractories, oxygen and carbon in the gas phase. Finally, we propose that hot Jupiters with silicate clouds and low water abundances might have accreted their envelopes between the soot line and the water snowline.

\end{abstract}

\section{Introduction}

Observations of both the solar system gas giants and extrasolar gas giants suggest that many have migrated away from their initial formation locations \citep[see reviews:][]{Morbidelli2018, Dawson2018, Fortney2021}. Since the composition of dust and gas varies with location in a protoplanetary disk, we expect the envelope compositions of gas giant exoplanets to encode information about where they formed \citep{Oberg2011}. For nearly a decade, this effort has focused on measurements of water and carbon bearing molecules such as carbon monoxide, which have strong spectral absorption bands at infrared wavelengths and are accessible to a variety of ground- and space-based telescopes \citep{Kreidberg2015, Wakeford2018, Welbanks2019, Madhusudhan2019}. This has in turn led most modeling studies to focus on predicting the enrichment of \textit{volatile} carbon and oxygen in giant planet atmospheres \citep[e.g.,][]{Madhusudhan2012, Espinoza2017}. Although this framework provides useful constraints on the potential formation locations of gas giant exoplanets, carbon and oxygen measurements alone are not enough to determine where a given gas giant planet originated \citep{Mordasini2016}.

We can break these degeneracies by characterizing the abundances of refractory elements (e.g., iron, silicon), which condense to form clouds at high temperatures, in exoplanet atmospheres. This is because the refractories are invariably condensed into the solid phase throughout most of the protoplanetary disk. In contrast, volatiles such as carbon and oxygen are delivered through the accretion of \htwoo, CO, and \cotwo ices, organics inherited from the interstellar medium (ISM), and are also present in disk gas. The volatile content of the accreted solids (in the form of ice or organics) therefore depends on where they condensed \citep{Oberg2011} and the extent of their thermal and chemical processing \citep{Lichtenberg2019, Lichtenberg2021}. By measuring the refractory abundance in a planet’s atmosphere, we can determine the fractional amount of solids incorporated into its envelope independent of where the solids originated \citep[e.g.,][]{Turrini2021}. The ratio of the refractory content relative to the volatile content in the atmosphere then provides us with a direct insight into the nature (formation location and degree of processing, which is related to size) of the solids that were incorporated into the planetary envelope \citep{Lothringer2021, Turrini2021, Schneider2021b}.

By measuring the refractory content of gas giant atmospheres, we can also improve our understanding of atmospheric carbon and oxygen. Refractory elements are important binders of oxygen \citep{Lodders2003}. This means that if we only measure the abundances of CO and \htwoo, we will systematically under-estimate the inventory of atmospheric oxygen. Measurements of the refractory content can also be used to quantify the relative amounts of carbon that were accreted in solid versus gas phase. This is important, because we currently have an incomplete understanding of the refractory carbon abundance of solids in the protoplanetary disk. In the ISM, half of the carbon reservoir is thought to be in the solid phase \citep{Mishra2015}. How much of this refractory carbon reservoir stays in solid phase in protoplanetary disks is a topic of active research. The refractory carbon abundance in solids plays a critical role in setting the overall carbon content of the atmospheric envelope \citep[e.g.,][]{Mordasini2016, Cridland2019, Turrini2021}, especially if the planet formed significantly interior to the \cotwo and CO snowlines.

\begin{figure*}
    \centering
    \includegraphics[width=\linewidth]{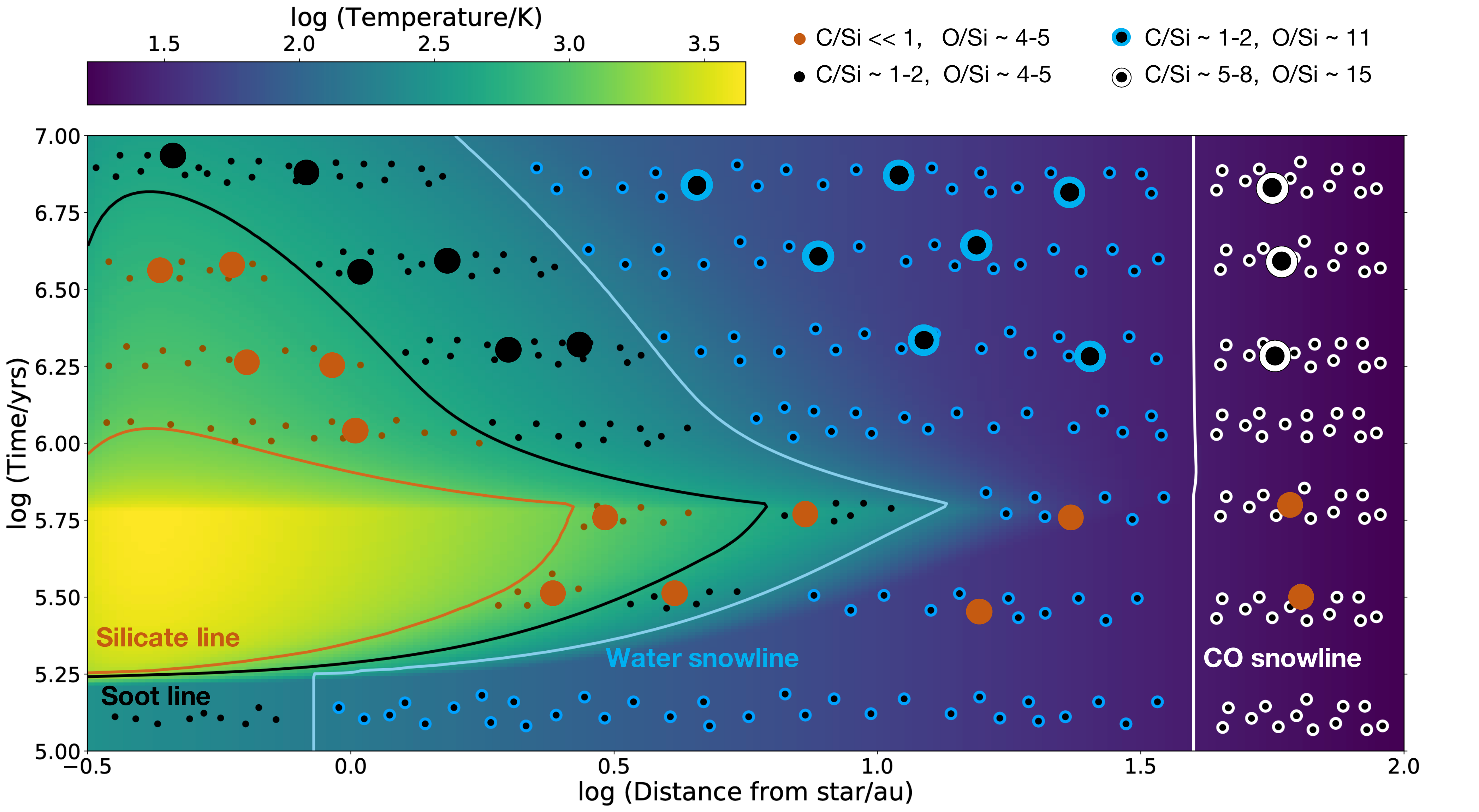}
    \caption{The evolution of a protoplanetary disk's temperature as a function of location and time based on the models presented in \cite{Drazkowska2018}. We overplot the condensation/sublimation lines of silicates (brown), `soots' (refractory carbon; black), water (blue), and CO (white) as solid lines. The expected composition of the small solids in each region is indicated in the legend on the top right. Even in the outer disk, planetesimals that form early might be volatile-poor if radionuclides are present. We represent these early-forming planetesimals as large brown circles (Si-rich composition). Later-forming planetesimals (large colored circles towards the top) will have compositions that reflect those of the solids in the local disk region.}
    \label{fig:disk_cartoon}
\end{figure*}

In this paper, we introduce a new framework for interpreting the atmospheric abundances of gas giant exoplanets. We define the atmospheric composition using three ratios: Si/H, O/Si, and C/Si. The Si/H ratio reflects the solid-to-gas ratio of the material incorporated into the envelope and is loosely equivalent to earlier notions of `metallicity'. It also stands in for the abundances of other refractory species that are expected to be accreted in solid phase \citep[e.g., see][who choose sulfur as a proxy for the refractory elements]{Turrini2021}. Instead of the more conventional C/O ratio, we use O/Si and C/Si. This alleviates the well-known degeneracies in the interpretation of atmospheric C/O ratios, and makes it easier to relate the measured abundance ratios to the compositions of the solids in the disk. In \S~\ref{sec:ref_comp}, we describe the expected spatial and temporal variation of O/Si and C/Si. For refractory C, which is particularly complicated, we merge constraints from the solar system meteorites with the most recent models for thermal processing within the protoplanetary disk. In \S~\ref{sec:formation_regimes}, we discuss our model assumptions and caveats, and then go on to predict the planetary envelope compositions associated with different formation regions. In \S~\ref{sec:uhjs}, we demonstrate that ultra-hot Jupiters are ideal targets for measuring all three elemental abundance ratios as all these elements are present in gas phase, providing new insights into their formation histories. In \S~\ref{sec:hj_comp}, we discuss how this framework can also be used to interpret the atmospheric compositions of hot Jupiters where silicates and other refractory species have already condensed. In \S~\ref{sec:conc}, we summarize our conclusions and discuss future work.

\section{Radially varying composition of solids and gas in the disk}
\label{sec:ref_comp}

\subsection{Refractory content of solids}
In this study, we assume that abundances of all refractory elements scale together for simplicity, i.e. the refractory elements are condensed out and only present in disk solids. The condensation region for the most abundant refractory materials such as silicon, iron, and magnesium is located very close to the star for ages greater than $\sim1$ Myr (silicate condensation line in Figure~\ref{fig:disk_cartoon}). We do not expect gas giant planets to accrete their gas envelopes this close-in, or at such early times \citep{Fortney2021, Drazkowska2022}. Therefore, the assumption that refractory elements are present only in solids is a valid first order assumption for studying the envelope composition of gas giant planets. This enables us to connect refractory abundance to a solid-to-gas accretion ratio for a planet.

Refractory elements have a range of `volatility' and the condensation sequence in the hotter regions of the disk will control each element's abundance in solids that condense out at different temperatures. Partial condensation can fractionate the major elements in disk solids and lead to variations in the abundance ratio of refractory elements in protoplanets. Based on meteoritic data from the solar system \citep{Wasson1988}, we might expect the abundance ratios of the most abundant refractories (magnesium, silicon, iron) to vary by a few tens of \% but not by an order of magnitude (e.g., Mg/Si varies from $\sim 0.7$ for enstatite chondrites to $\sim 1.1$ for carbonaceous chondrites; Mg/Si $\sim 1.2$ for the bulk silicate Earth, \citealt{McDonough1995}). Our assumption that the abundances of these refractories scale together is therefore reasonable. 

However, the abundance ratio of refractory species with higher volatilities could deviate significantly from solar proportions. Moderately volatile refractory elements such as sodium, potassium, and sulfur can be significantly fractionated relative to elements such as silicon in protoplanets due to partial condensation in the nebula or heating-induced volatile loss in planetesimals. The measured abundances of these moderately volatile elements can therefore correspond to a lower limit on the solid-to-gas accretion rate. Elements such as sulfur, which may be primarily present in solids in protoplanetary disks \citep{Kama2019, McClure2020}, may provide an opportunity to trace the refractory compositions of planets too cool to retain highly refractory elements such as silicon in gas-phase \citep{Turrini2021}. Nonetheless, since it is unclear how refractory the solid form of sulfur is and whether it is in an unidentified gas phase reservoir \citep{LeGal2021} as well as the complicated nature of sulfur chemistry in hot Jupiter atmospheres \citep{Hobbs2021, Polman2022}, highly refractory species such as silicon offer a better tracer of solid accretion.

High resolution observations of ultra-hot Jupiters provide us with an opportunity to measure relative abundances of refractory elements \citep[e.g.,][]{Gibson2022}. Recent ground-based observations of WASP-76b, for example, show that many refractory elements in its atmosphere are present in solar proportions relative to iron with the exception of titanium (likely condensed out), sodium, and potassium \citep{Pelletier2022}. Sodium and potassium are highly depleted relative to iron, most likely due to ionization at the extremely hot temperatures in the planet's atmosphere. 

For the rest of this study, we will use silicon to represent the ensemble of all refractory species, as it is one of the most abundant refractory elements in protoplanetary disks \citep{Asplund2009}. Moreover, silicon is commonly used as an anchor in abundance measurements of the Sun and meteorites in our solar system \citep[e.g.,][]{Bergin2015, Lodders2019}. Using silicon as a proxy for the overall refractory abundance therefore allows us to more easily connect our study with the wider solar system literature. We note, however, that a measured atmospheric abundance for any refractory element can be used to constrain the solid-to-gas accretion ratio of the planet's gas envelope in our framework.

\subsection{O/Si content of solids}
\label{sec:OtoSi_content}

We begin by characterizing the oxygen abundance relative to silicon in solids. For this, we need to know how oxygen is distributed in its various forms. In the ISM, measurements indicate that 25\% of oxygen is tied to refractory elements, assuming a solar oxygen abundance \citep{Whittet2010}. Since solar O/Si $\sim 15.1$ \citep{Asplund2009} this would imply O/Si $\sim 4$ in refractory solids, which condense interior to the water snowline. This number is in agreement with the inferred stoichiometric ratio of silicon and oxygen in the ISM solids \citep{Draine2003}. However, 40\% of oxygen in the ISM is missing from the current inventory (Unidentified Depleted Oxygen or UDO). It is suspected that hidden water ice might account for a large part of this missing oxygen \citep{Neill2013}. An issue with this explanation is that in warm shocks, the water never reaches the abundance needed to hide the oxygen in this way \citep{Leurini2015, Kristensen2017}. Another possibility is that oxygen is present in some moderately refractory phase (e.g., polyoxymethylene in comets) that only volatilizes at a few hundred K \citep{vanDishoeck2021}. If half of the `missing' oxygen (UDO) is somehow tied to such refractories, the O/Si of solids within the water snowline would be $\sim 7$ \citep[e.g., see][]{Turrini2021}. \cite{Lodders2019} suggest that hydrated silicates and/or hydrous sulfates could store a large amount of water and lead to O/Si $\sim 7$. However, most hydrated silicates have O/Si $\lesssim 4.5$\footnote{Out of common phyllosilicates, 1:1 type phyllosilicates, such as kaolinite and serpentine, will typically contain O:Si of 9:2 \citep{Brady2008}. Common 2:1 di-octahedral phyllosilicates typically have a base O:S1 of 12:4 \citep{Brady2008} but have the capacity to take up a maximum of 3.6 \htwoo during enhanced swelling, typically in response to high relative humidity \citep{Ferrage2005}. As such, phyllosilicates would rarely have a higher O/Si than 4.5.}, so any additional refractory oxygen might be tied to carbon in some organic or polymeric form rather than to metal oxides; the former would be converted into the gas phase within the soot line (see \S~\ref{sec:CtoSi_content}).

In this work we assume that O/Si $\sim 4$ in refractory solids, corresponding to a scenario in which the missing oxygen is primarily tied up in water. This means that in our models, silicon and other refractories are highly enriched relative to oxygen in the solids interior to the water snowline. Beyond the water snowline, water ice condenses out and transfers a large fraction of the oxygen budget to the solid phase. This increases the solid O/Si in this region to $\sim 11$. Farther out in the disk, the condensation of \cotwo and CO causes the oxygen content of solids to increase until their O/Si ratio reaches the solar value of $15.1$. Figure 1 shows a simplified sketch of the location- and time- dependent O/Si content of disk solids.

\begin{table*}
	\begin{threeparttable}
    	\caption{Adopted elemental ratios for solids and gas in the disk} \label{table:adopted_comp}
    	\begin{center}
        	\begin{tabular*}{\textwidth}{@{\extracolsep{\fill}} l l c c c c c c}
        		\hline \hline
        		Scenario & Location & \multicolumn{2}{c}{O/Si (solids)}  & O/H (gas) & \multicolumn{2}{c}{C/Si (solids)} & C/H (gas) \\
        		 & & Elemental ratio & $\times$ solar & $\times$ solar & Elemental ratio & $\times$ solar & $\times$ solar \\ \hline 
        		1 & Within soot line & 3.8 & 0.25 & 0.75 & 0 & 0 & 1  \\
        		2 & Between soot line \& \htwoo snowline & 3.8 & 0.25 & 0.75 & 1 & 0.12 & 0.88 \\
                3 & Just within \htwoo snowline\textsuperscript{a} & 3.8 & 0.25 & 10.75 & 1 & 0.12 & 0.88 \\
                4 & Just beyond \htwoo snowline\textsuperscript{a} & 162.4 & 10.73 & 0.27 & 1 & 0.12 & 0.88 \\
                5 & Between \htwoo and CO snowlines & 11.1 & 0.73 & 0.27 & 1 & 0.12 & 0.88  \\
                6 & Just within CO snowline\textsuperscript{a} & 11.1 & 0.73 & 5.77 & 1 & 0.12  & 10.88  \\
                7 & Just beyond CO snowline\textsuperscript{a} & 98.3 & 6.50 & 0 & 91.5 & 11 & 0 \\
                8 & Beyond CO snowline & 15.14 & 1 & 0 & 8.32 & 1 & 0  \\
        		\hline
        	\end{tabular*}
        	 \begin{tablenotes}
             \small
			 \item {\bf Notes.} 
			 \item \textsuperscript{a}{Assuming $10 \times$ solar enrichment of \htwoo or CO in the gas phase (within the snowlines) or solid phase (outside the snowline).}
			\end{tablenotes}
    	\end{center}
	\end{threeparttable}
\end{table*}

\subsection{C/Si content of solids}
\label{sec:CtoSi_content}

In the outermost regions of the disk, the carbon abundance in solids due to condensation of CO and \cotwo is straightforward to calculate. This picture becomes more complicated interior to the \cotwo snowline, where the carbon budget of the solids is determined by the abundance of refractory carbon. In the ISM, observations indicate that half of the total carbon budget is locked up in the refractory phase, primarily as refractory organics \citep{Mishra2015, Oberg2021}. However, a significantly lower refractory carbon abundance is measured in meteorites, sun-grazing comets, and in the bulk content of the terrestrial planets (e.g., C/Si $\sim 10^{-3}$ for the Bulk Silicate Earth) in the solar system. The under-abundance of carbon in these materials relative to the ISM suggests that refractory carbon must be depleted in the inner disk \citep{Bergin2015}. Even the most pristine and least processed carbonaceous meteorites exhibit some refractory carbon depletion \citep{Wasson1988}, indicating that these depletion processes must occur on relatively small bodies and at relatively low temperatures.

Some studies have argued that the destruction of refractory carbon may be ascribed to oxidation or photolysis of small grains in the disk surface layers \citep{Bauer1997, Finocchi1997, Gail2002, Lee2010, Alata2014, Alata2015, Visser2007}, but this would require efficient vertical mixing \citep{Anderson2017} and extreme parameter combinations in order to match the solar system observations \citep{Klarmann2018}. It instead seems more likely that thermal processing of the grains inherited by the protoplanetary disk causes them to irreversibly lose their refractory carbon once they are exposed to temperatures $\gtrsim 500$ K, forming the so called `soot' line \citep{Kress2010, Li2021}. This sublimation of refractory carbon leads to the production of volatile gas phase carbon species that do not condense in the inner disk \citep{Kress2010, Wei2019}, hence the irreversibility. In this framework, we would expect to see a decreasing abundance of refractory carbon as we move towards the inner disk \citep[e.g.,][]{Gail2017, Dartois2018} with grains within the soot line almost entirely devoid of refractory carbon. This picture is roughly supported by the solar system data on the composition of meteorites and comets \citep[e.g.,][]{Bergin2015, Woodward2021}.

In this study, we assume that solids that are exposed to temperatures greater than 500 K lose their refractory carbon entirely \citep[C/Si $<< 1$][]{Li2021}. For solids exterior to the soot line and interior to the \cotwo snowline, we set the refractory carbon abundance equal to that of the least processed and most pristine meteorites from our solar system, which have C/Si $\lesssim 1$ (we choose the upper bound of C/Si = 1). It is possible that the abundance of refractory carbon in solids might be even higher in more distant regions of the disk that were never exposed to high temperatures and where solids did not undergo appreciable thermal processing (e.g., both \citealt{Cridland2019} and \citealt{Turrini2021} assume a higher refractory carbon abundance). In order to limit the number of distinct compositional scenarios in the discussion section, we assume a constant abundance between the soot line and the CO snowline in our semi-quantitative model. We also assume that the CO and \cotwo snowlines are co-located; this is justifiable in the context of our models because \cotwo constitutes a relatively small reservoir of carbon and oxygen (see \citealt{Eistrup2016, Eistrup2018} for a possibly larger role for \cotwo). Beyond the CO snowline, the condensation of carbon bearing volatiles increases the C/Si in solids to $\sim 5 - 8$.

In our work, we have assumed that the composition of the circumstellar disk is inherited from the parent molecular cloud and used the ISM composition to set the initial oxygen and carbon repositories of solids and gas in the disk (although, for refractory carbon we have incorporated some effects of irreversible loss). This assumption is supported by the composition of comets and the isotopic composition of water in the solar system as well as observations of protoplanetary disks \citep[e.g.,][]{Cleeves2014, Drozdovskaya2019, Booth2021, vanDishoeck2021}. However, the refractory composition of some meteorites and terrestrial planets in the solar system suggest that infalling material was completely vaporized and recondensed as the disk cooled \citep[][`chemical reset']{Grossman1972, Davis2006}. It is likely that both chemical reset and inheritance from protostellar cloud are at play in protoplanetary disks \citep{Pontoppidan2014, Oberg2021}. Chemical reset has a significant influence on the distribution of oxygen and carbon in the circumstellar disk \citep{Eistrup2016, Eistrup2018}. We point the reader to \cite{Cridland2019} and \cite{Pacetti2022} for a more detailed description of how such a chemical reset would affect the composition of materials accreted by giant planets.

\subsection{The effect of disk evolution on composition}

As giant planet formation is an integrative process, we need to know the temperature of the disk as a function of distance and time in order to prescribe the composition of solids and gas. Our temperature structure is take from the models of \cite{Drazkowska2018}, who use a simple infall model in combination with viscous disk evolution. We do not explicitly incorporate the effect of dust and gas evolution on their composition. Instead, we assume that the solid and gas carbon and oxygen repositories add up to solar values at each location before accounting for pebble drift and evaporation.

The radial drift of solids is expected to bring volatile material inward. This material sublimates at the snowlines and can locally enhance the abundance of carbon and oxygen rich molecules \citep{Oberg2016}. The sublimated gas then evolves viscously and diffuses radially. This is predicted to enhance the abundance of carbon and/or oxygen interior to the relevant snowline \citep[e.g.,][]{Booth2017, Booth2019, CevallosSoto2022}. Just beyond the snowline, the outward diffusion of this gas can lead to recondensation \citep{Stevenson1988, Schoonenberg2017, Drazkowska2017}, enhancing the relative abundance of carbon and/or oxygen in the icy solids. In locations affected by such processes, the abundances of these elements in gas and solid phase can become much larger than solar values. The accretion of enriched gas and solids from these regions can be an important source of heavy element enrichment in planetary envelopes \citep[e.g.,][]{Schneider2021a, Bitsch2022}. We note that although refractory carbon is also expected to sublimate at the soot line, the resulting carbon rich gas does not recondense when it diffuses beyond the soot line because of the irreversibility of refractory carbon sublimation.  This means that this process only produces an enrichment of carbon in the gas phase near the soot line.

To study the effect of pebble drift and evaporation heuristically, we calculate the composition of a planet near the snowline assuming a $10 \times$ solar enrichment of \htwoo (CO) in gas phase just interior to the \htwoo (CO) snowline and the same enrichment in solid phase just beyond the \htwoo (CO) snowline. Previous studies show that these are reasonable values to adopt here \citep[e.g.,][]{Schoonenberg2017, Booth2017}. The exact degree of enrichment and the extent of the disk region that is enriched are a function of the pebble flux (rate at which solids cross the snowlines), the strength of the radial diffusion and viscous evolution, as well as the presence of drift barriers in the outer regions of the disk \citep{Booth2017, Kalyaan2021}. Moreover, we note that these effects of pebble evaporation and recondensation would be weak in disks in which most of the pebble mass has been converted into planetesimals. Current \citep{Banzatti2020} and upcoming observations (e.g., JWST GO 1640 \& 2025) should provide the first empirical constraints on the pebble evaporation driven enrichment of volatile gases interior to snowlines in the inner disk.

\subsection{Accretion and processing of planetesimals}

So far, we have discussed the solid composition of dust ($\mu$m sized grains to cm sized pebbles) present in the disk and the effect of evaporation and condensation on the dust's oxygen and carbon content. In the pebble accretion framework, protoplanets grow by accreting these pebbles, rapidly forming massive planetary cores outside several au \citep{Ormel2010, Lambrechts2012}. However, once these growing planetary cores become sufficiently massive, they perturb their surrounding gas disk in a way that cuts off the flow of pebbles onto the planet \citep{Morbidelli2012, Lambrechts2014a}. From this point on, the accretion of larger solid planetesimals ($\sim 1 - 100$ km) is expected to become one of the dominant sources of heavy element enrichment in the outermost layers of gas giant envelopes \citep[e.g.,][]{Alibert2018, Venturini2020a}. These planetesimals are thought to form from pebbles and/or smaller dust grains in their local region of the protoplanetary disk via processes such as the streaming instability \citep{Youdin2005, Johansen2007}. It is therefore reasonable to expect that the initial compositions of these planetesimals will reflect those of the smaller solids they formed from.

However, unlike smaller solids, these planetesimals might lose a significant fraction of their initial volatile content as a result of degassing and devolatilization due to heating from short-lived radionuclides such as $^{26}$Al \citep{Grimm1993, Monteux2018}. The importance of radionuclide heating for thermal processing of planetesimals depends on the the disk's initial $^{26}$Al abundance, the timing of planetesimal formation, and the size of the planetesimals \citep[e.g.,][]{Lichtenberg2016}. Different stellar systems are expected to form with a wide range of initial $^{26}$Al abundances as this radionuclide is thought to originate from a supernova in the molecular cloud's/nascent star's vicinity \citep{Lichtenberg2016a, Lugaro2018}. In disks with a significant $^{26}$Al reservoir, planetesimals that form early reach high internal temperatures, which causes them to lose their volatiles. However, due to the short half-life of $^{26}$Al (0.7 Myr), planetesimals that form late only undergo significant thermal processing if they are large ($\gtrsim 100$ km). We do not directly model these effects in this work, but we do comment on the consequences of this effect where it is relevant.

\section{Final compositions of gas giant envelopes: breaking degeneracies with Si/H, O/Si, and C/Si}
\label{sec:formation_regimes}

\subsection{Model assumptions and caveats}
We are required to make a number of assumptions in order to relate the compositions of the solids and gas in the disk to the final compositions of planetary atmospheres. We summarize these assumptions here and elaborate on the various caveats associated with our choices. Forming planets likely accrete solids and gas and migrate through the protoplanetary disk simultaneously. As they grow, planets may migrate through multiple compositionally distinct regions and their final compositions can therefore be a mixture of the materials from adjacent regions \citep{Cridland2019, Turrini2021, Khorshid2021}. However, the extent to which gas giant planets migrate during their formation is poorly constrained by current observational studies, and theoretical estimates can vary by many orders of magnitude \citep{Armitage2010}. 

It seems unlikely that the population of close-in giant planets universally begin forming at a few tens of au (a location at which it is difficult to form massive cores) and then migrated all the way to the inner disk. Large scale migration models tend to overestimate the occurrence rates of hot Jupiters \citep{Ida2008, Bitsch2019, Emsenhuber2021} and also disagree with observations of the solar system \citep{Deienno2022}. Moreover, radial velocity surveys indicate that giant planet occurrence peaks at intermediate distances ($\sim 1 - 10$ au, \citealt{Fulton2021}). It is difficult to produce such a peak in models with large scale migration \citep{Hallatt2020}, suggesting that many of these gas giants may have originated in this region \citep[e.g.,][]{Lambrechts2014a,  Chachan2021, Emsenhuber2021}. 

In light of these considerations, we elect to focus on scenarios in which the final envelope compositions of giant planets are inherited from a single compositional region. It is relatively straightforward to modify these predictions to reflect scenarios where the planets accreted from multiple compositionally distinct regions, as the final envelope composition would then simply be a linear combination of these regions. However, such a procedure is bound to increase the non-uniqueness of our inferences about a planet's formation region. We note that the fractional time a planet spends accreting in each region will also depend on the spatial extent of the region, which can vary significantly ($\sim 1$ au between soot line and \htwoo snowline vs a few tens of au between the \htwoo and CO snowlines, see Figure~\ref{fig:disk_cartoon}).

It is also worth noting that when we predict the envelope compositions of giant planets, we only need to explain the composition of the outer well-mixed convective envelope. Observations of Jupiter and Saturn \citep{Wahl2017, Mankovich2021} suggest that the interiors of giant exoplanets may also be compositionally stratified, either as a result of their accretion histories or subsequent core erosion \citep{Helled2017, Helled2022}. Although the outer convective layer is expected to move deeper into the planet with time \citep{Vazan2015}, the composition of the underlying, stably stratified regions is effectively separated from that of the uppermost, observationally accessible layers of the atmosphere. This effectively reduces the window of time in which the planet can move through compositionally distinct regions of the disk and still have an effect on the final observed atmospheric abundances. It also means that our observations of giant planet envelopes cannot probe the earlier stages (e.g., core formation) of their formation history, which could have happened at a different location.

In this work, we are also agnostic about the mechanism by which solids are accreted and the extent of the heavy element enrichment that is possible in different regions of the disk. The accretion of solids during the runaway phase of gas accretion is poorly understood and current formation models have difficulty explaining the high metal contents of giant planets both in the solar system and beyond \citep[e.g.,][]{Thorngren2016, Shibata2019}. Planets are expected to stop accreting pebbles once they become massive enough to trap dust beyond their orbit \citep{Morbidelli2012, Lambrechts2014a}.  At this point, planetesimal accretion is thought to become the dominant mode of solid accretion. 

\begin{figure*}
    \centering
    \includegraphics[width=\linewidth]{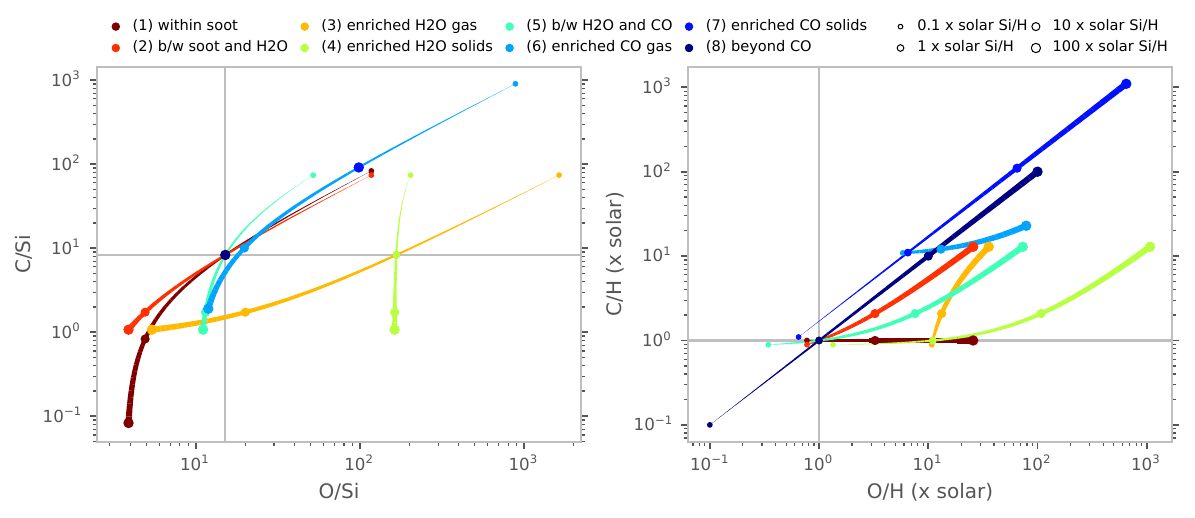}
    \caption{Predicted abundance ratios for different formation regions (see legend on the top). These ratios depend on the total amount of solids accreted relative to the gas, as well as the composition of the solids. The Si/H ratio, which parameterizes the solid-to-gas accretion ratio, is indicated by the line thickness. Si/H values of 0.1, 1, 10, and 100 $\times$ solar are marked with filled circles of increasing size. The solar values for each elemental ratio are indicated as thin grey lines for reference.}
    \label{fig:elem_ratio}
\end{figure*}

However, dynamical models predict that the planetesimal accretion rate will decrease as the planet becomes more massive and excites the orbits of nearby planetesimals \citep{Zhou2007, Shiraishi2008, Shibata2019, Eriksson2022}. Giant planets can accrete more planetesimals if the planetesimal surface density is higher than the typically assumed Minimum-Mass-Solar-Nebula \citep{Dodson-Robinson2009, Venturini2020a} or if they migrate over tens of au \citep{Shibata2020, Knierim2022}, but as discussed earlier, it is unclear whether such large scale migration happens in most planetary systems. It has also been suggested that planetary mergers from giant impacts at large distances ($\sim 10$ au) could enrich giant planets in heavy elements \citep{Ginzburg2020}. Given the uncertainty surrounding the processes by which giant planets accrete solids, we do not place any limits on the allowed values for Si/H in our compositional models, where Si/H is a measure of the amounts of solids accreted relative to gas.

In this study we define Si/H, C/H, and O/H relative to the solar value for convenience. However, stellar hosts span a range of elemental abundances and the ratios of these elements show some correlations with stellar metallicity, with [O/Si] and [C/Si] varying by as much as $\sim 0.5$ dex in a given stellar metallicity range \citep[e.g.,][]{Ness2018, Buder2018}. When applied to individual exoplanetary systems, our model predictions should therefore be rescaled to account for the measured composition of the host star \citep[e.g.,][]{Bitsch2020, Turrini2021, Pacetti2022}. However, we note that in our discussion of atmospheric chemistry in \S~\ref{sec:hj_comp}, it is the number ratios of the relevant elements that determine the equilibrium molecular composition of the atmosphere. Although, these results do not depend on the normalization with respect to the star, the stellar abundance ratios do determine the elemental number ratios that can be attained by a planet forming in different compositional regions.

\subsection{Compositions inherited from different formation regions}

Now that we have defined the compositions of the solids and gas in each region of the disk, we can predict the resulting elemental ratios in the envelopes of gas giant planets as a function of their formation location. For each location we must assume a value for the Si/H ratio, which parameterizes the ratio of solids to gas incorporated into the planetary envelope. The C/Si and O/Si ratios in a planet's atmosphere are then calculated as a function of the assumed Si/H ratio for each location by our semi-quantitative model. We give these values in Table~\ref{table:SitoH_vary} and plot the results in Figure~\ref{fig:elem_ratio}.

\begin{table*}
	\begin{threeparttable}
    	\caption{Elemental ratios for different scenarios with varying Si/H} \label{table:SitoH_vary}
    	\begin{center}
        	\begin{tabular*}{\textwidth}{@{\extracolsep{\fill}}l l | c c c c | c c c c}
        		\hline \hline
        		Scenario & Location & \multicolumn{4}{c |}{O/Si} & \multicolumn{4}{c}{C/Si} \\ \hline
        		 & \hspace{2.5cm} Si/H ($\times$ solar) & 0.1 & 1 & 10 & 100 & 0.1 & 1 & 10 & 100\\ \hline 
        		1 & Within soot line & 117.30 & 15.14 & 4.92 & 3.90 & 83.18 & 8.32 & 0.83 & 0.08 \\
        		2 & Between soot line \& \htwoo snowline & 117.30 & 15.14 & 4.92 & 3.90 & 74.18 & 8.32 & 1.73 & 1.07 \\
                3 & Just within \htwoo snowline\textsuperscript{a} & 1630.86 & 166.49 & 20.05 & 5.41 & 74.18 & 8.32 & 1.73 & 1.07 \\
                4 & Just beyond \htwoo snowline\textsuperscript{a} & 203.27 & 166.49 & 162.81 & 162.45 & 74.18 & 8.32 & 1.73 & 1.07 \\
                5 & Between \htwoo and CO snowlines & 51.92 & 15.14 & 11.46 & 11.09 & 74.18 & 8.32 & 1.73 & 1.07 \\
                6 & Just within CO snowline\textsuperscript{a} & 883.68 & 98.31 & 19.78 & 11.92 & 905.94 & 91.49 & 10.05 & 1.90  \\
                7 & Just beyond CO snowline\textsuperscript{a} & 98.31 & 98.31 & 98.31 & 98.31 & 91.49 & 91.49 & 91.49 & 91.49 \\
                8 & Beyond CO snowline & 15.14 & 15.14 & 15.14 & 15.14 & 8.32 & 8.32 & 8.32 & 8.32  \\
        		\hline
        	\end{tabular*}
        	 \begin{tablenotes}
             \small
			 \item {\bf Notes.} 
			 \item \textsuperscript{a}{Assuming $10 \times$ solar enrichment of \htwoo or CO in the gas phase (within the snowlines) or solid phase (outside the snowline).}
			\end{tablenotes}
    	\end{center}
	\end{threeparttable}
\end{table*}

Inside the soot line (scenario 1), our model predicts that the solids are highly depleted in carbon with an O/Si $\sim 4$. Planets with high Si/H that accrete their envelopes in this region should thus have a very low carbon abundance (C/Si $<< 1$) in their atmospheres. So far, we have not found any close-in gas giant planets that are rich in oxygen and poor in carbon (i.e. with C/O $<<1$). Between the soot line and the water snowline (scenario 2), the solids have an O/Si $\sim 4$ and C/Si $\lesssim 1$. Planetary envelopes with a high Si/H ($\gtrsim 10 \times$ solar) that form in this region should have Si-rich atmospheres where carbon and silicon act as the primary binders of oxygen. If the temperature of the atmosphere is cool enough for silicates to condense, this has important implications for the planet's water abundance. We discuss this in more detail in \S~\ref{sec:hj_comp}. 

Just within the \htwoo snowline (scenario 3), the gas phase oxygen abundance is enhanced by $10 \times$ solar in our fiducial model. Planets with low Si/H that form here would have $5-10 \times$ higher oxygen abundances than those that form further in from the \htwoo snowline. For Si/H $\gtrsim 100 \times$ solar, the disk solids dominate the composition of the planetary envelope and the enhanced oxygen in the gas becomes negligible. Similarly, just beyond the \htwoo snowline (scenario 4), the water ice to refractory mass ratio of solids is enhanced relative to regions further out in the disk. Formation in this region has interesting implications for how oxygen-rich and carbon-poor (and consequently water-rich) a planet can be. For such planets, O/Si remains highly super-solar and C/Si becomes sub-solar with increasing Si/H. Moving further out, formation between the \htwoo and CO snowlines (scenario 5) is well studied and produces planets with high C/O if they have low Si/H $< 1 \times$ solar and sub-solar C/O if solid accretion dominates (Si/H $> 1 \times$ solar). 

Just within and beyond the CO snowline (scenario 6 \& 7), the same considerations apply for the enrichment of CO, in the gas and solid phase respectively, as discussed earlier for the region around \htwoo snowline. The main difference is that the accretion of CO enriched gas or solids leads to an enrichment of both carbon and oxygen. The enrichment of CO in gas phase just within the CO snowline (scenario 6) has drawn particular interest as formation in this region is currently considered the only way a giant planet can possess a high C/O \textit{and} super-solar C/H and O/H \citep[e.g.,][]{Oberg2016}. In \S~\ref{sec:hj_comp}, we show that there is another formation pathway that can produce planets with such compositions at $T \lesssim 2000$ K. In fact, CO enriched solids just beyond the snowline (scenario 7) also present an interesting compositional scenario that can produce planets with high O/Si and C/Si and C/O close to 1 for any Si/H. Further out beyond the CO snowline (scenario 8), nearly the entire carbon and oxygen budget is condensed in the solids. As a result, we expect the C/Si and O/Si to be solar and constant with Si/H for planets forming in this region. 

These distinct formation locations correspond to fairly different spaces in C/Si vs O/Si (as well as C/H and O/H) \textit{for a given} Si/H. Although simultaneous migration and accretion might lead to a mixture of material from two or more of the compositionally distinct regions studied here, this framework can be easily adapted to encompass such scenarios. Even within this relatively simple framework, our results demonstrate that measuring Si/H, O/Si, and C/Si simultaneously is a promising way to reduce the degeneracies inherent in our interpretations of planetary envelope compositions.

\section{Ultra-hot Jupiters are an Ideal Population for Measuring Refractory Abundances}
\label{sec:uhjs}

\begin{figure*}
    \centering
    \includegraphics[width=\linewidth]{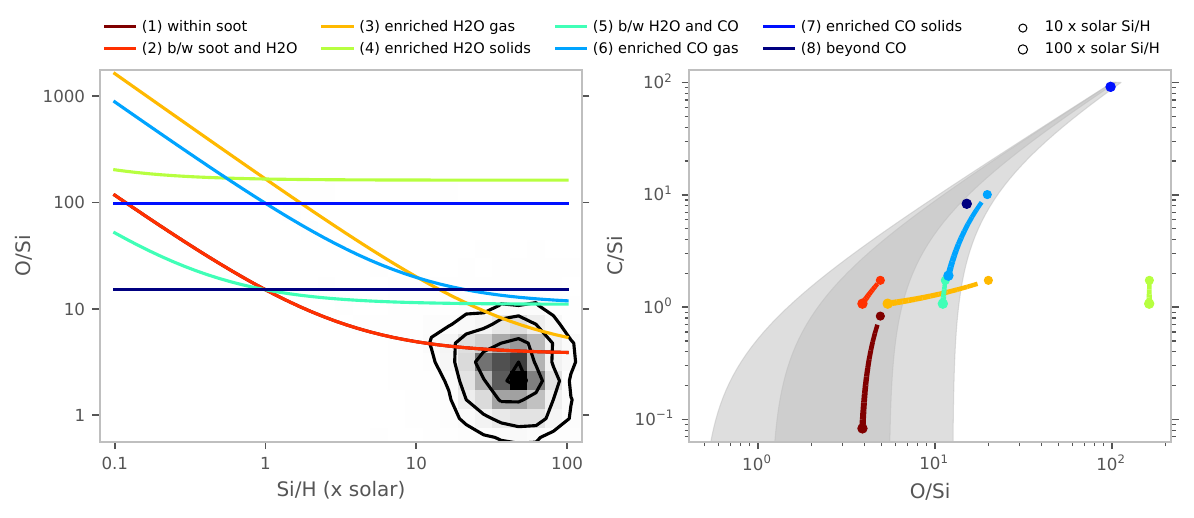}
    \caption{In the left panel, we show the constraints on O/Si and Si/H for WASP-121b obtained in \cite{Lothringer2021} in grey, and compare them to model predictions (assuming solar values for O/Si and C/Si) for each of the different regions discussed in \S~\ref{sec:formation_regimes}. These measurements are best-matched by a model that is enriched in refractories relative to oxygen, suggesting that this planet finished forming interior to the water snowline. However, the oxygen abundance reported in this study does not include any contribution from CO, and therefore is likely an under-estimate. In the right panel, we vary the CO abundance and show the corresponding C/Si as well as its effect on the 1 and 2 $\sigma$ O/Si constraints obtained \cite{Lothringer2021}. Upcoming IR measurements with \emph{HST} and \emph{JWST} will measure the CO abundance for this planet, placing tighter constraints on its formation location.}
    \label{fig:wasp121b_inferences}
\end{figure*}

The same property that makes silicon and other refractory elements ideal for tracking solid accretion (namely, the fact that they condense at relatively high temperatures) also make them challenging to measure in planetary atmospheres. Fortunately, a subset of the most highly irradiated gas giant planets (known as `ultra-hot Jupiters') have atmospheric temperatures that are high enough ($T_{\rm eq} \gtrsim 2000$ K) to retain their refractories in gas phase \citep{Kitzmann2018, Lothringer2018, Parmentier2018}. Their high atmospheric temperatures also make them some of the most favorable exoplanetary targets for transmission spectroscopy, as they have relatively large atmospheric scale heights and little to no aerosol opacity \citep{Evans2018, Fu2021, Helling2021a, Wilson2021}. 

Refractory elements have especially prominent absorption features in the near-UV and optical bandpasses \citep{Lothringer2020} and many of them have now been detected in ultra-hot Jupiter atmospheres using a combination of ground-based high resolution spectroscopy \citep[e.g.,][]{Gaudi2017, Hoeijmakers2018, Cabot2020, Ehrenreich2020, Gibson2020, Merritt2021} and space-based low to medium resolution spectroscopy \citep[e.g.,][]{Sing2019, Lothringer2022}. Although detections of refractory species in ultra-hot Jupiter atmospheres abound, only a handful of planets (e.g., WASP-121b, \citealt{Lothringer2021}; WASP-178b, \citealt{Lothringer2022}; WASP-76b, unpublished as of now, \citealt{Pelletier2022}; KELT-9b, \citealt{Kasper2021}; KELT-20b, \citealt{Kasper2022}) have robust statistical constraints on the absolute abundances of their atmospheric refractory content.

WASP-121b's transmission spectrum has been measured from $0.3 - 1.7 \; \mu$m \citep{Evans2018, Sing2019}, which allowed \cite{Lothringer2021} to place constraints on both the planet's water and refractory content. In Figure~\ref{fig:wasp121b_inferences}, we show the Si/H and O/Si for WASP-121b measured by \cite{Lothringer2021} in the context of our framework. We note that we have used solar values of O/Si and C/Si for these models. A proper comparison would require using the composition of WASP-121 to calculate the model predictions. We find that the scenarios most compatible with these observations corresponds to envelope accretion between the soot line and the water snowline. Our ability to rule out other scenarios is limited by the lack of a constraint on WASP-121b's carbon content as we do not even have the planet's transit depths in the \emph{Spitzer} bandpasses. Since carbon is an important binder of oxygen in the form of CO, we may be systematically under-estimating the total amount of oxygen when we calculate the O/Si ratio using only the measured water abundance. 

We can resolve this degeneracy by measuring the CO abundance and corresponding C/Si ratio in the planet's atmosphere. If the planet accreted its envelope between the soot line and the \htwoo snowline, we would expect to measure a small ($\lesssim 1$) C/Si ratio. If the measured CO abundance and corresponding C/Si ratio is higher, our measured value for the O/Si ratio will increase as well. This moves the posterior higher up in Figure~\ref{fig:wasp121b_inferences}. In the right panel of this figure, we show the effect of CO abundance variation on the inferred O/Si for WASP-121b. For now, we can only rule out the accretion of solids that were enriched in \htwoo ice just beyond the water snowline (scenario 4). Upcoming \emph{JWST} observations (GO 1201 \& 1729) will measure this planet's CO abundance, effectively resolving the current degeneracies in the interpretation of its measured SiO and \htwoo abundances.

WASP-178b is also a particularly promising target for the characterization of Si/H, O/Si, and C/Si. \cite{Lothringer2022} published the NUV and optical ($0.2 - 0.8 \; \mu$m) spectrum of WASP-178b with \emph{HST} UVIS and observed a dramatic rise in transit depth ($\sim 20$ scale heights) shortward of $0.4 \, \mu$m that is well matched by absorption from SiO in the planet's atmosphere. During the coming year, WASP-178b will be observed in the $0.8 - 1.7 \; \mu$m bandpass with \emph{HST} (GO 16086 \& 16450) and $3 - 5 \; \mu$m bandpass with \emph{JWST} (GO 2055). When combined with the NUV and optical data, these observations will provide tight constraints on the carbon, oxygen, and silicon abundances in this planet's atmosphere. Our models demonstrate that this will in turn place strong constraints on the composition of the accreted solids and the solid-to-gas accretion ratio of WASP-178b's envelope. There are several other ultra-hot Jupiters that are comparably favorable targets, but we need observations of more planets to characterize them at a population level and leverage their unique properties to develop a better understanding of gas giant formation.

\subsection{Caveats in translating gas phase molecular abundances to bulk elemental ratios}

There are some caveats that are worth keeping in mind when fitting refractory absorption features in order to infer elemental abundance ratios. Firstly, close in planets are tidally locked and as a result have strong day-night temperature contrasts. The temperature at which a refractory species condenses out is typically significantly lower (a few hundred K) than the equilibrium temperature of a tidally locked planet for which we no longer see the refractory species in gas phase. This is due to condensation and sequestration (rain out) of the refractory species on the night side, which is significantly cooler than the planet's equilibrium temperature \citep{Lothringer2020}. Determining whether rain out is a gradual or a sudden process is therefore crucial for correctly inferring refractory abundances in the gas phase. In addition, 3D effects can bias retrieved abundances for these atmospheres, which have large temperature gradients \citep{Pluriel2020, Pluriel2022, MacDonald2020, Taylor2020, Welbanks2021}. Observational characterization of these process with transit observations in the near-UV and in the optical with high resolution that target refractory species would be valuable for this making headway on this problem.

It is also important to ensure that all the major bearers of a particular refractory element are accounted for while calculating its elemental abundance. For example, extremely high temperatures on the day sides of ultra-hot Jupiters likely lead to silicon being present in both atomic and molecular (SiO) form \citep[e.g.,][]{Lothringer2018, Lothringer2020a, Helling2021}. Measuring its abundance therefore requires some knowledge of the local conditions (e.g., temperature, degree of dissociation) of the observed atmosphere. Finally, unresolved absorption lines from escaping refractories may be confused with strong continuum absorption features in low resolution observations. It is therefore critical to establish the origin of the spectral features before making abundance inferences. For example, \cite{Lothringer2022} observed WASP-178b in high resolution with \emph{HST} STIS to resolve any absorption from escaping magnesium and/or iron and did not detect these species, which allowed them to infer that their low resolution NUV spectrum was tracing continuum absorption from SiO.

\section{The origin of water-poor hot Jupiters}
\label{sec:hj_comp}

As we move from ultra-hot Jupiters ($T \gtrsim 2000$ K) to more conventional hot Jupiters ($1000 \gtrsim T < \lesssim 2000$ K), we can draw on more than a decade of atmospheric observations from space- and ground- based observatories for atmospheric characterization studies \citep[see][for a review]{Madhusudhan2019}. However, we expect that the most common refractory elements, such as iron and silicon, will have condensed out into clouds in these atmospheres \citep[e.g.,][]{Marley2015, Gao2021}. For these planets, mid-infrared observations targeting absorption features of silicate clouds combined with optical spectra that constrain cloud particle size distribution may allow us to put some constraints on the cloud mass and consequently the refractory abundance \citep{Wakeford2015, Min2020}. However, even with perfect knowledge of the particle size distribution, the morphology (amorphous versus crystalline grains) and aggregate nature of cloud particles will introduce additional uncertainty in the inferred cloud mass.

Moderately volatile refractory species such as sodium (Na) and potassium (K) might also be useful probes of refractory content in hot Jupiter atmospheres, as they persist in the gas phase down to temperatures as low as $\sim 1000$ K \citep{Lodders1999, Burrows2001}. Sodium and potassium abundances have been estimated for many hot Jupiter atmospheres \citep[e.g.,][]{Sing2016, Nikolov2018, Welbanks2019}. As noted earlier, these moderately volatile refractory species can be used as tracers of the solid-to-gas accretion rate as long as the accreted solids are not significantly depleted in them. For accreted solids that formed at high temperatures with fractionated compositions due to partial condensation, the measured Na and K abundances provide us with a lower limit on the solid-to-gas accretion rate.

Alternatively, we might infer the presence of abundant condensed refractories in hot Jupiter atmospheres by leveraging their effect on the gas-phase oxygen abundance. When refractory species such as silicon and magnesium are in gas phase, they are typically present in the form of SiO and Mg. However, when they condense, they are converted to enstatite or forsterite (MgSiO$_3$ or Mg$_2$SiO$_4$), which binds additional oxygen atoms to silicon. These additional oxygen atoms are taken from \htwoo present in the gas, which reduces its abundance in the atmosphere. However, the CO abundance remains unchanged as oxygen is tightly bound to carbon (preferred by Gibbs free energy minimization). This increases the apparent C/O ratio in the gas phase in a planet's atmosphere \citep{Woitke2018, Helling2021a}.

\begin{figure}
    \centering
    \includegraphics[width=\linewidth]{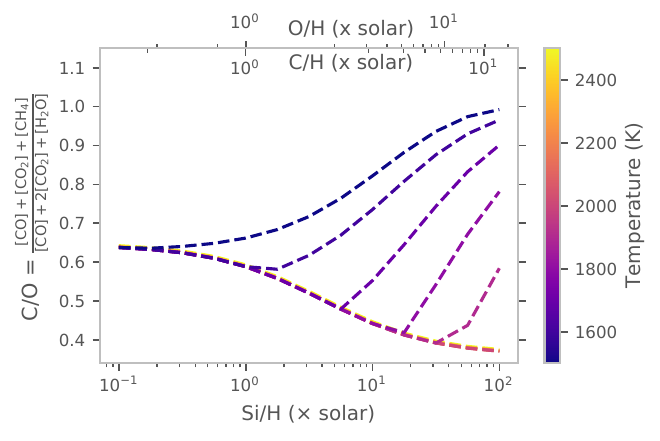}
    \caption{The C/O of a planet measured using only CO, \cotwo, \htwoo, and CH$_4$ plotted against Si/H for a planet forming between the soot line and the water snowline (disk solids with O/Si = 3.8 and C/Si = 1). Once the composition becomes dominated by solid accretion, i.e. the planetary atmosphere's O/Si and C/Si approach that of the accreted solids, the measured C/O becomes temperature dependent and for a given Si/H, rises sharply with decreasing temperature. This is due to condensation of silicates, which removes a large fraction of oxygen from \htwoo.}
    \label{fig:ctoo_condense}
\end{figure}

If silicon and oxygen are present in solar proportions this effect is relatively small, with condensation only removing $\lesssim 25$\% of the oxygen. However, as we note in our study, silicon and oxygen do not always have to be present in solar proportions. In fact, for the formation regimes discussed here, silicon is almost always expected to be enriched by a greater amount than oxygen for atmospheres with high solid-to-gas accretion ratios (e.g., high Si/H ratios). This holds everywhere except beyond the CO snowline and perhaps just beyond the \htwoo snowline, where grains might be enriched in ice due to condensation of outward diffusing water vapor.

Planets that accrete their envelopes within the water snowline can have O/Si ratios as low as $\sim 4 - 5$ (compared to solar value of 15.1). For planets with these low O/Si ratios, the condensation of silicates has a much larger effect on the gas phase oxygen abundance. In Figure~\ref{fig:ctoo_condense}, we use a simple chemical setup with GGChem \citep{Woitke2018} that includes only H, He, C, O, Si, Fe, and Mg to show this effect. For planets forming between the soot line and the \htwoo snowline, we show how the gas phase C/O (as measured by the typically detected species CO, \htwoo, \cotwo, and CH$_4$) at a pressure of 0.1 bar depends on temperature. At high temperatures the measured C/O decreases with Si/H, which reflects the relatively carbon-poor composition of solids in this region. However, the measured C/O rises for cooler temperatures due to the condensation of silicates and this effect is more pronounced at higher Si/H because it corresponds to lower O/Si (Table~\ref{table:SitoH_vary} and Figure~\ref{fig:elem_ratio}).

Published low resolution studies of hot Jupiter atmospheres have reported tentative evidence for subsolar \htwoo abundances \citep{Barstow2017, Pinhas2019, Welbanks2019}. This result was recently confirmed by high resolution spectroscopy of several planets, which show that subsolar \htwoo is often accompanied with supersolar CO abundances (\citealt{Giacobbe2021}; VMR log(\htwoo) $< -5.66$ and VMR log(CO) = $-2.46^{+0.25}_{-0.29}$ for $\tau$ Boo b, \citealt{Pelletier2021}; VMR log(\htwoo) = $-4.4 \pm 0.4$ for HD 189733 b, \citealt{Boucher2021}). Consequently, the measured C/O ratios of these planetary atmospheres are very high ($0.85 - 1$). In Figure~\ref{fig:ctoo_condense}, we demonstrate that formation between the soot line and the \htwoo snowline can produce C/O ratios that are consistent with values reported in these studies for a sufficiently high Si/H ($\gtrsim 10 \times$ solar). 

In published formation models, these C/O and C/H ratios can only occur if these planets accreted gas enriched in CO due to evaporating pebbles close to the CO snowline \citep[e.g.,][]{Oberg2016, Booth2017}. Our new proposed formation pathway for water depleted hot Jupiters has several advantages over formation near the CO snowline. It significantly reduces the distance that the planets must migrate over, from $20 - 30$ au for the CO snowline to $\sim$several au for formation between the soot and \htwoo snow lines (this can vary depending on the timing of envelope accretion, see Figure~\ref{fig:disk_cartoon}). In this scenario, hot Jupiters also originate in the inner part of the region where giant planet occurrence rate peaks \citep[$\sim 1 - 10$ au, ][]{Fernandes2019, Fulton2021}, providing a consistent overarching picture of giant planet formation. 

The simplest way to differentiate between the established hypothesis and our proposed scenario is to measure the refractory content of the planetary envelope. If these planets attained a high C/O and supersolar C/H by accreting CO rich gas, they should have accreted a relatively small quantity of solids (Si/H $\lesssim 1 \times$ solar). In contrast, our proposed scenario requires Si/H $\gtrsim 10 \times$ solar. Using sodium and potassium as tracers of refractory content, it should be possible to distinguish between these two hypotheses. Indeed, the tentatively observed higher enrichment of sodium and/or potassium relative to water in hot Jupiters \citep{Welbanks2019} appears to align with the prediction from our hypothesis. Recently, \cite{Hands2022} showed that a `fully formed' giant planet migrating from 5 au to within the water snowline can accrete enough planetesimals within the snowline to be enriched in sodium and potassium while remaining water-poor. However, they do not account for oxygen inherited from the accretion of these planetesimals or during the prior formation process, even though in their model formation occurs primarily beyond the water snowline. In the future, it would be useful to combine our framework with a planetesimal accretion model like the one presented in their work to see if planetesimal accretion can provide the enrichment required to explain the observations.

If hot Jupiters and ultra-hot Jupiters all accreted their envelopes in the same region of the disk, we can also differentiate between our proposed and the canonical scenarios by characterizing the gas-phase carbon and oxygen abundances in these atmospheres as a function of temperature. If these planets accreted their envelopes between the soot and \htwoo snowlines, we would only expect to see low \htwoo abundances for planets that are cool enough for silicates to condense. If most close-in gas giants instead originated near the CO snowline, ultra-hot Jupiters would also have a high C/O and low water abundance even though their atmospheres are too hot for silicates to condense. If we can obtain a large sample of measurements of water abundance or C/O (by measuring abundances of common molecules such as CO, \htwoo. \cotwo, and CH$_4$) as a function of atmospheric temperature, we should be able to differentiate between these two proposed hypotheses.

We note that the validity of our proposed formation pathway for water-poor hot Jupiters depends on the C/Si and O/Si of solids and gas between the soot line and the water snowline. Two important factors that affect these elemental ratios are: i) variations in the disk's C/Si and O/Si due to differences in stellar composition, and ii) a different refractory oxygen and carbon abundance than assumed in our study. As for the variations in disk and stellar compositions, stars with higher metallicities tend to have a lower O/Si and C/Si and a higher C/O than the Sun \citep{Ness2018, Buder2018, Bitsch2020}. This trend makes it easier to attain the conditions necessary for producing the required level of water-depletion due to silicate condensation (i.e. a higher atmospheric C/O at $T \lesssim 2000$ K for a lower Si/H compared to Figure~\ref{fig:ctoo_condense}). The high metallicities of hot Jupiter host stars \citep{Santos2004, Fischer2005, Fulton2021} therefore render our proposed scenario for the planets' water-depleted compositions more plausible. Detailed compositional characterization of the planet-hosting stars \citep[e.g.,][]{Reggiani2022, Kolecki2022, Biazzo2022} would provide a route to test our hypothesis.

As noted in \S~\ref{sec:ref_comp}, the refractory oxygen and carbon abundances beyond the soot line could be higher \citep{Oberg2021, vanDishoeck2021, Turrini2021}. Our proposed scenario for hot Jupiter water depletion remains plausible for higher carbon and oxygen refractory abundances as long as the solids have C/Si + $x~$(Si/Si) $\sim$ O/Si, where $x \sim 3 $ depending on which silicates condense out \citep{Visscher2010, Gao2021}. This is because we only need enough carbon and silicon (i.e., metal) atoms to bind up all of the available oxygen in order to deplete the atmospheric water abundance. With a higher C/O value beyond the soot line, solids with C/Si and O/Si that satisfy this condition can help reach the required level of atmospheric water depletion more easily than in our baseline scenario. More observations of protoplanetary disks that inform us about the distribution of oxygen and carbon are therefore critical for making these inferences and connecting formation with composition.

\section{Future Work and Conclusions}
\label{sec:conc}

In this work, we have shown that measuring refractory elemental abundances significantly enhances our ability to relate the envelope compositions of gas giant planets to their formation locations and accretion histories. Refractory elements can be used to constrain the planet's solid-to-gas accretion ratio, resolving degeneracies in the interpretation of complementary carbon and oxygen abundances. This in turn allows us to determine the relative partitioning of accreted volatile elements between the gas and solid phases. This information uniquely constrains the region(s) in the disk where the planet accreted its gas envelope and solids. Importantly, we show that the current widely-adopted retrieval framework in which one scales either the carbon or oxygen abundance along with the refractories using a global metallicity knob is erroneous \citep[see also][]{Turrini2021, Pacetti2022}. Carbon and oxygen abundances should only scale with the refractory abundance beyond the CO snowline, where all the heavy elements are condensed out.

We use our model framework to predict how the measured Si/H, O/Si, and C/Si abundances in giant planet envelopes vary as a function of the assumed solid-to-gas ratio and accretion location. These predictions can be directly compared to literature measurements of elemental abundances in the atmospheres of gas giant exoplanets, keeping in mind that a planet may have accreted from more than one compositionally distinct region.  Ultra-hot Jupiters are some of the most promising targets for these studies, as almost all the elements (including the refractories) are present in gas phase. We use WASP-121b and WASP-178b as examples to demonstrate how upcoming \textit{JWST} and \textit{HST} transmission spectroscopy of these objects can be combined with published observations to provide the first simultaneous constraints on the abundances of silicon, carbon, and oxygen. Complementary studies of ultra-hot Neptunes such as LTT 9779b \citep{Jenkins2020} and TOI-849b \citep{Armstrong2020} may prove even more interesting, as these planets are predicted to exhibit a wider range of Si/H (solid-to-gas accretion) ratios than their Jovian counterparts \citep{Fortney2013}.

Using this new framework, we also identify a previously unrecognized pathway for forming water-poor hot Jupiters, which had been thought to originate near the CO snowline.  We show that these planets might instead have primarily accreted their well-mixed outer envelopes in the region between the soot line and the \htwoo snowline; both theoretical and observational studies suggest that this is a more favorable region for giant planet formation than the CO snowline. Planets forming in this region accrete solids that are relatively poor in oxygen (O/Si $\sim 4$) and rich in carbon (C/Si $\sim 1$). If the atmosphere of the hot Jupiter is cool enough for silicates to condense, this will remove a large fraction of the atmosphere's oxygen, which in turn reduces the gas-phase water abundance.  We argue that future measurements of potassium and/or sodium abundances in these atmospheres could be used to differentiate between these two scenarios.

If we wish to expand the sample of gas giant planets with measured refractory abundances, there are several potential approaches we might utilize.  Many refractory elements have prominent absorption lines at near-UV and optical wavelengths, which can be detected using transmission spectroscopy. Atomic iron and SiO (for ultra-hot planets) and sodium and potassium (for slightly cooler planets) are readily observable at these wavelengths. Although there are multiple published detections of refractory species using high resolution transmission spectroscopy \citep[e.g.,][]{Borsa2021, Merritt2021, Bello-Arufe2022, Kesseli2022, Prinoth2022}, relatively few of these studies use retrievals to constrain their abundances.  Recently developed techniques for carrying out atmospheric retrievals using high resolution data sets \citep{Brogi2019, Gibson2020, Fisher2020} should be applied to the plethora of published observations of ultra-hot Jupiter atmospheres that report the detection of refractory species.

There may also be additional opportunities to characterize refractory abundances at infrared wavelengths. For example, SiO has prominent absorption bands near $\sim 4 \; \mu$m (L band) and $8 \; \mu$m \citep{Barton2013}. It is difficult to detect these bands in hydrogen-rich atmospheres using low resolution spectroscopy, because water tends to dominate the opacity at these wavelengths unless SiO is more abundant (although unlikely, this is not impossible). However, it may be possible to target this molecule using high resolution observations in \emph{L} band, which could resolve the distinct absorption lines of these two molecules. These studies could be carried out using either emission or transmission spectroscopy.

There are complementary approaches to the framework presented in this work that also enable us to thread a connection between formation and composition. For cooler planets ($\lesssim 500$ K), \emph{JWST} is expected to detect nitrogen bearing molecules for the first time.  Like carbon, the nitrogen content of disk solids is a complex function of distance and time and depends on the poorly known refractory nitrogen repository. However, a majority of the nitrogen is much harder to detect than carbon in molecular clouds and protoplanetary disks, most likely because it is present in the form of the homonuclear diatomic molecule N$_2$. It is thus typically assumed that a large fraction of the nitrogen repository is volatile interior to the NH$_3$ and N$_2$ snowlines. However, refractory organics and ammonium salts could constitute a significant fraction of the total nitrogen content, as indicated by observations of solar system comets \citep{Altwegg2020, Oberg2021}. Measurement of the nitrogen abundance in exoplanet atmospheres could therefore be an important supplementary probe of the accreted solids, although disequilibrium chemistry at cooler temperatures might affect the interpretation of these measurements \citep[e.g.,][]{Moses2014}. Several published studies have already begun to explore how nitrogen abundances in giant planet atmospheres might be used to constrain their formation histories \citep{Cridland2020, Turrini2021, Pacetti2022}. It would be straightforward to incorporate nitrogen into our model framework in a future study. 

The measurement of isotopologues in planetary atmospheres also presents a potential way of determining whether volatile elements such as carbon were primarily accreted via gas or solids \citep{Molliere2019, Morley2019}. The recent measurements of CO isotopologue ratios in two brown dwarf atmospheres deviate moderately from the expected ISM value, with one of them hinting at the enrichment of $^{13}$C likely due to accretion of fractionated CO ice \citep{Zhang2021b, Zhang2021a}. However, a more robust connection between formation and isotopologue composition needs to be established before we can fully interpret these measurements.

In addition to these observational studies, there is also more work that could be done to improve our framework for interpreting these abundances. Although we have made reasonable semi-quantitative assumptions about the time variation of the solid and gas composition of the disk, we could further improve the accuracy of our predictions by incorporating a more quantitative disk model that properly accounts for the dynamics of dust and gas \citep[e.g.,][]{Oberg2016, Booth2017, Cridland2019}. Doing so would allow us to relax some of our current assumptions, e.g. that the gas and solid components of each element add up to the solar abundance at a given location. Uniting this framework with a planetesimal formation and evolution model would allow us to track the composition of larger solids that might be incorporated into the planetary envelope \citep[e.g.,][]{Lichtenberg2021}. For planets forming in the inner disk, it is also important to explore the effect of radial drift barriers on the evolution of disk solid and gas compositions in this region.  Finally, a retrieval setup for our composition and formation framework \citep[e.g.,][]{Molliere2022} would be useful for interpreting upcoming \emph{JWST}, \emph{HST}, and ground based observations and for making inferences about formation histories of gas giant exoplanets. All these tasks are highly tractable and worthwhile in light of the rapidly growing body of observational data.

\acknowledgments
We thank the referee for a thoughtful and detailed report that helped us improve our paper. Y.C. would like to thank Aida Behmard and Eva Linghan Scheller for enlightening discussions about the variations in stellar composition and the hydration of silicates respectively. Y.C. acknowledges partial support from the Natural Sciences and Engineering Research Council of Canada (NSERC) through the CITA National Fellowship and the McGill Space Institute through the MSI Fellowship. H.A.K. would like to acknowledge support from NASA/STScI through a grant linked to the HST-GO-14767 program.

\software{\texttt{astropy} \citep{astropy2018}, \texttt{GGChem} \citep{Woitke2018}}

\bibliography{manuscript}
\bibliographystyle{apj}

\end{document}